\documentclass[%
 aip,
 amsmath,amssymb,
 reprint,%
]{revtex4-1}

\usepackage{graphicx}
\usepackage{dcolumn}
\usepackage{bm}

\usepackage[utf8]{inputenc}
\usepackage[T1]{fontenc}
\usepackage{mathptmx}
\usepackage{etoolbox}
\usepackage{xcolor}%

\makeatletter
\def\@email#1#2{%
 \endgroup
 \patchcmd{\titleblock@produce}
  {\frontmatter@RRAPformat}
  {\frontmatter@RRAPformat{\produce@RRAP{*#1\href{mailto:#2}{#2}}}\frontmatter@RRAPformat}
  {}{}
}%
\makeatother
\begin{document}

\preprint{AIP/123-QED}

\title{Conditions for Suppression of Gas Phase Chemical Reactions inside a Dark Infrared Cavity: O$_{2}$ + 2NO → 2NO$_{2}$ as an Example}
\author{Mwdansar Banuary}
 \affiliation{Schulich Faculty of Chemistry, Technion-Israel Institute of Technology, Haifa, 32000, Israel}
\author{Ashish Kumar Gupta}%
\affiliation{ 
Department of Chemistry, Indian Institute of Technology Guwahati, Assam, 781039, India
}%

\author{Nimrod Moiseyev}
\email{nimrod@technion.ac.il}
 \homepage{https://nhqm.net.technion.ac.il/}
\affiliation{%
Schulich Faculty of Chemistry, Technion-Israel Institute of Technology, Haifa, 32000, Israel
}%
\affiliation{Faculty of Physics, Technion-Israel Institute of Technology, Haifa, 32000, Israel}

\date{\today}

\begin{abstract}
The ability to slow down chemical reactions using a seemingly simple setup—reactions confined within a cavity formed by two parallel mirrors—is fascinating. However, theory and experiment have not yet fully converged.
In this work, we provide the conditions and guidelines for selecting reactions that can be suppressed in a dark cavity. The primary requirement is that the reaction's potential energy surface must contain two saddle points ({and not one saddle point as required for enhancement}). This condition enables a reduction in the reaction rate. Specifically, we demonstrate that the reaction rate of O$_{2}$ + 2NO$\to$ 2NO$_{2}$  can be suppressed by a dark cavity composed of two parallel mirrors. We show that the suppression of the reaction rate depends on the distance between the mirrors, which determines the cavity parameters, and on the number of molecules in the transition state configuration that simultaneously interact with the cavity. 
\end{abstract}

\maketitle

\section{\label{Intro}Introduction} 
The application of a dark cavity made of two mirrors to control reaction rates has attracted researchers in recent years. In 2012, Thomas Ebbesen and his co-workers\cite{hutchison2012modifying} showed for the first time that the rate of a chemical reaction can be suppressed by vacuum fields. This research explores the interactions between light and matter at the quantum level and their implications for chemical processes, which is known as polaritonic chemistry. Most recently, in 2024, it was shown  that the rates of asymmetric reaction with a \textit{single} saddle point in the potential energy surface can be enhanced by a dark cavity\cite{moiseyev2024conditions}. In this work, we show that the reaction having two \textit{saddle points} can be suppressed by a dark cavity.

There are numerous theoretical efforts exploring the mechanisms underlying polaritonic chemistry. Several reviews have discussed about the change of the chemical reactions by polaritons (see references \cite{feist2018polaritonic,hertzog2019strong,herrera2020molecular,li2022molecular,ebbesen2023introduction,bhuyan2023rise,mandal2023theoretical,campos2023swinging,weight2025ab}).  In Refs. \cite{ben2025enhanced,felicetti2020photoprotecting}, the non-Hermitian formalism of quantum mechanics (NHQM) was applied to study the cavity control of photochemical reactions. In this work, we follow this line, but instead focus on the effect of dark cavity on chemical reactions. Here we take an approach where the complex decay poles of the scattering matrix, so called Gamow-Siegert resonances, are used as a basis set in the calculations of the rates of chemical reactions in cold molecular collisions in gas phase. This approach has previously been applied for studying the conditions for enhancement of the rate of reactions in a dark cavity  (no photon is inserted into the cavity) \cite{moiseyev2024conditions}. As we will show here the conditions for suppression of chemical reaction's rates are different. 

 In this work, we use the reaction path Hamiltonian approach where the dynamics in multidimensional space is described by one dimension which is the curvilinear coordinate that is considered as the reaction coordinate. For example, in a collinear scattering experiment of the reaction $A + BC \to AB + C$, the reaction coordinate $r_{RC}$ describes the transition from the reactants $A + BC$, where $BC$ is in its rovibrational ground state, to the products $AB + C$, where $BC$ is also in its rovibrational ground state, via a transition state that corresponds to a saddle point on the two-dimensional potential energy surface (2D PES). In this 2D PES, the $X$-axis represents the distance between atom $A$ and the center of mass of the $BC$ molecule, while the $Y$-axis corresponds to the internuclear distance within the $BC$ molecule. The reaction coordinate is defined such that as $r_{RC} \to -\infty$, $X \to \infty$ and $Y \to R_{eq}^{BC}$, the equilibrium bond length of $BC$. When $r_{RC} = 0$, the $X$ and $Y$ coordinates define the geometry of the $ABC$ activated complex at the transition state (TS), which is a saddle point on the 2D PES. As $r_{RC} \to +\infty$, the products $AB + C$ are obtained, with $BC$ in its rovibrational ground state. { In our case, the 2D PES contains two saddle points. The potential along the reaction coordinate therefore forms a well between two asymmetric barriers, with entrance and exit channels leading to the intermediate region. Outside the cavity, the relative kinetic energy of the reactants is sufficient for the reaction to proceed without tunneling through the barriers. Upon collision, the activated complex is formed in a metastable state with an energy approximately equal to the maximum height of the barriers.  }

 Here, we included calculations of the rate when all molecules interact among themselves via the cavity. This calculations can be done because we use only two resonances for each molecule as basis functions. One resonance we use as a basis function is the transition state (TS) resonance which is associated with the shape type excited resonance (third-node in our case) which is located at the top of the potential barrier as obtained from the 1D calculations along the curve linear reaction coordinate system that provides the rate of the reaction outside of the cavity.  The second resonance function is the 0-node narrowest resonance  (so called the ground state resonance, GR state) which is localized at the bottom of the potential well in between the two double barrier potential along the curve linear reaction coordinate. 
 
 { The polaritons (hybrids of molecular and radiation states) are the eigenfunctions of the Hamiltonian, which will be introduced later. This Hamiltonian describes a system of $N$ molecules inside the cavity, where the molecules interact only through their coupling to the cavity field. The polaritons are calculated as linear combinations of polaritonic basis functions: $|Polariton\rangle_\alpha =\Sigma_{n-photon=0}^N C_{n-photon,\alpha}|Polariton-basis\rangle_{n-photon}$. }

Let us describe several so called  polariton basis functions. $$|Polariton-basis\rangle_{0-photon}=\Pi_{j=1}^N |TS\rangle_{j-mol}|0_{photon}\rangle_{j-mol}$$ for $N$ molecules in the vacuum of the QED field.  { Notice that the TS is an excited molecular state (with three nodes in our case) of the activated complex. It arises from the relative kinetic energy of the reactants, which is sufficient to overcome the reaction barrier outside the cavity, even in the photon vacuum state (i.e., in the absence of photons).}

\begin{eqnarray}
  &&  |Polariton-basis\rangle_{1-photon}=\\ \nonumber &&
    \Pi_{j=1;j'\ne j}^N|GR\rangle_{j-mol}|1_{photon}\rangle_{j-mol}|TS\rangle_{j'-mol}|0_{photon}\rangle_{j'-mol}
    \end{eqnarray}
    
    for one photon emission from any one of the N molecules. The emitted photon energy is $\hbar\omega_{cav}=Re[E_{res}^{TS}-E_{res}^{GR}]$

{ To make clear our notation notice that $|TS\rangle_{j-mol}$ implies that $j-th$ molecule is its TS state (3 node excited resonance), and $|GR\rangle_{j-mol}$ is when the $j-th$ molecule is in its ground state (no node) resonance state.}

\begin{widetext}
\begin{equation}
|Polariton-basis\rangle_{2-photon}= \Pi_{j=1;j'\ne j; k\ne j; k\ne j'}^N|GR\rangle_{j-mol} |1_{photon}\rangle_{j-mol} |GR\rangle_{j'-mol} |1_{photon}\rangle_{j'-mol}|TS\rangle_{k-mol}|0_{photon}\rangle_{k-mol}
\end{equation}
\end{widetext}

for two photon emissions from two different molecules.

{ 
For the  case of polaritonic state which is a mixed state of one photon and N molecules then $N-1$ molecules are in the ground state and only one in rovibrational excited state. Therefore, this polaritinic state is a linear combination of N components where different molecule is excited. When the polaritonic state is of two photons then the molecular function is linear combination of all possible configurations where N-2  molecules are in the ground state and only two molecules are rovibrational excited. 
}
 Since the number of polariton basis functions is growing fast with increasing N we could not consider more than 10 molecules inside the cavity.

 \textit {The important significant result of our calculations is that the effect of the cavity on suppression of the rate of the reaction is observable also when the density of molecules which interact only with the cavity is increased. See Fig.\ref{pol00} that shows the ratio between the reaction rate as obtained by using the polariton basis set given above with the reaction rate as obtained outside of the cavity when the distance between the two mirrors is infinite large. }

{Let us explain why this result may seem surprising. It contradicts the usual situation in which all 
$N$ molecules are initially in the ground state (rather than in the excited TS resonance, as in our case). In that case, their interaction with a single photon in the cavity leads to an average reaction rate equal to the rate obtained outside the cavity. Notice that when the number of molecules 
$N$ inside the cavity increases, the molecular density in the cavity also increases.}

Here, the NHQM will be utilized to calculate the complex poles, which has previously been applied in studying the cavity effects in chemistry. These studies discovered a novel mechanism for generating high-purity single-photon emission at high repetition rates\cite{ben2023non} and the conditions for upper polariton to penetrate into the continuum and to become a metastable state with a finite lifetime\cite{moiseyev2022polariton}. However, in the case of conventional quantum mechanics, in solving this kind of problem, it becomes challenging as one has to deal with a wavepacket rather than a single time independent eigenstate. NHQM enables the solution of this problem in a simple way which requires a solution of time-independent Schrodinger equation (TISE) imposing outgoing boundary conditions. The complex eigenvalues obtained from this solution are associated with the complex poles of the scattering matrix. In a chemical reaction, these complex poles describe the predissociation of metastable states of the activated complex.

\section{\label{sec2}Methods}

To study the reaction rates within a dark cavity, electronic energy as a function of the reaction coordinate and the harmonic frequencies perpendicular to the reaction coordinate are required. The use of one of the complex scaling methods discussed in ref.\cite{moiseyev2011non} is also required to calculate the complex poles which are used as a basis set to calculate reaction rates. In this work, the smooth exterior scaling (SES) method is applied.

The main condition for a reaction to be suppressed by a dark cavity is that the reaction must contain two saddle points in the potential energy surface. Therefore, the following reaction has been chosen to study the suppression of chemical reactions inside a dark microwave cavity.
\begin{equation}
    2NO + O_{2} \rightarrow 2NO_{2} \label{r1}
\end{equation}

\subsection{Reaction Path Hamiltonian from Potential Energy Surface}

This reaction had been proposed to proceed mainly through the following three open channels\cite{mckee1995ab,tsukahara1999gas,olson2002conformation,gadzhiev2009quantum,gadzhiev2011mechanism}.

(i) Two molecules of NO react with one O$_{2}$ molecule to produce 2NO$_{2}$. This reaction proceeds through an intermediate with two transition states.

\begin{equation}
    2NO+O_{2} \rightarrow TSa \rightarrow Intermediate
\end{equation}

\begin{equation}
    Intermediate \rightarrow TSb \rightarrow 2NO_{2}
\end{equation}

(ii) In this channel, a NO molecule is combined with an O$_{2}$ molecule in the first step, forming an intermediate, NO$_{3}$. The intermediate reacts with another NO molecule to produce the product.

\begin{equation}
    NO+O_{2} \rightarrow NO_{3}  \label{m1}
\end{equation}

\begin{equation}
    NO_{3} + NO \rightarrow 2NO_{2}  \label{m2}
\end{equation}

(iii) Another channel is the formation of a dimer of NO as an intermediate.

\begin{equation}
    NO+NO \rightarrow (NO)_{2} 
\end{equation}

\begin{equation}
    (NO)_{2} + O_{2} \rightarrow 2NO_{2}
\end{equation}

 Gadzhiev and co-workers\cite{gadzhiev2011mechanism} studied the mechanism (i) using the CCSD(T) and CASSCF methods. They reported that the localization of TSb in the CASSCF method was complicated by the intruder states. Using the CCSD(T) method too, the calculation of complete potential energy surface (PES) is difficult and computationally costly. It is also mentioned that the results obtained in ref. \cite{gadzhiev2009quantum} by DFT method are invalid. Tsukahara and coworkers\cite{tsukahara1999gas} stated that channels (ii) and (iii) are more favorable than channel (i), which requires three body collisions simultaneously. The complex (NO)$_{2}$ is difficult to describe accurately\cite{mckee1995ab}. Therefore, in this work, we choose the mechanism (ii), which is a two-step mechanism with the formation of NO$_{3}$ intermediate. The complete geometry optimization of the stationary points is done at the B3LYP/6-31G(d) level using Gaussian 16\cite{g16}. The minima are confirmed by obtaining no imaginary frequency, and the transition states are confirmed by obtaining exactly one imaginary frequency.

\subsubsection*{Step 1}
In Step 1 (reaction \ref{m1}), two transition states (TS1 and TS2) are obtained with an intermediate (INT). The intrinsic reaction coordinates (IRCs) for both TS1 and TS2 are calculated. Both IRCs go up to optimized points except towards the reactant. Some points are added from the relax scan of the distance between the center of mass of NO molecule and the center of mass of O$_{2}$ molecule following the Generalized Internal Coordinates (GIC) approach to the IRC of TS1 towards the reactant to connect to the optimized structure of the reactant. Connecting the IRCs and the points obtained from the relax scan, a complete potential energy curve for the step 1 reaction is obtained. From the coordinates of this curve, the following mass weighted coordinates are calculated.

(i) \textbf{Coordinate 1} $(x)$ is the distance between the center of mass of O$_{2}$ and the center of mass of NO which is multiplied by $\sqrt{\frac{\mu_{O_{2}-NO}}{m_{O}}}$, where $\frac{1}{\mu_{O_{2}-NO}}=\frac{1}{m_{O}+m_{O}}+\frac{1}{m_{N}+m_{O}}$

(ii) \textbf{Coordinate 2} $(y)$ is the bond length of $O_{2}$ multiplied by $\sqrt{\frac{\mu_{O_{2}}}{m_{O}}}$, where $\mu_{O_{2}}$ is the reduced mass of O$_{2}$. 


Using the \textbf{coordinate 1} and \textbf{coordinate 2}, another coordinate ($r$) corresponding to the electronic energies V(r) is calculated following the relation
\begin{equation}
   r(i+1)=r(i) - \sqrt{(x(i+1)-x(i))^{2}+(y(i+1)-y(i))^{2}} \label{coord_1}
\end{equation}
where r(1)=0.

\begin{figure}
\centering
  \includegraphics[width=\linewidth]{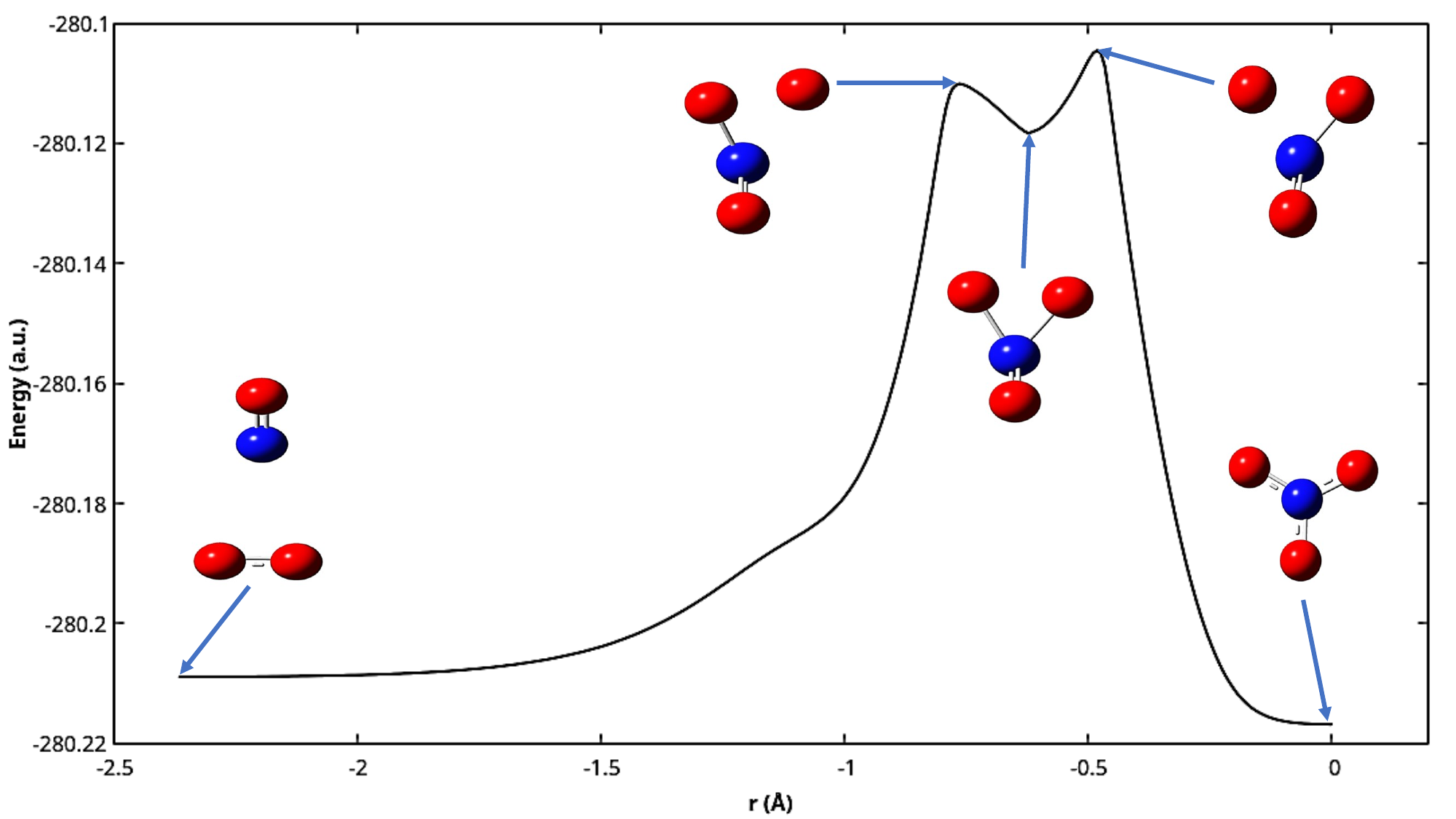}
  \caption{Potential Energy curve for step 1, O$_{2}$+NO $\rightarrow$ NO$_{3}$ reaction along with optimized structures at stationary points. $r$ is the mass weighted coordinate obtained from coordinate 1 and coordinate 2 by using the relation \ref{coord_1}. The molecular structures, which are shown by arrows from left to right, correspond to the optimized structures of reactant (O$_{2}$+NO), TS1, INT, TS2, and intermediate product (NO$_{3}$) at $r$=0. This NO$_{3}$ will react with another NO molecule to produce the final product (2NO$_{2}$).}
  \label{fig_4}
\end{figure}

Using the newly generated coordinate ($r$), the potential energy curve for Step 1 is obtained as given in Fig. \ref{fig_4}

\subsubsection*{Step 2} 
 In step 2 (reaction \ref{m2}), the NO$_{3}$ obtained in step 1 is combined with another NO molecule to produce the final product (2NO$_{2}$). The IRC for transition state (TS3) for this step is calculated, but it does not go to minima, unlike in the case of step 1. The minima are obtained by optimizing the last points of the IRC on both sides. Therefore, the new mass weighted \textbf{Coordinate 3} and \textbf{Coordinate 4} for the step 2 reaction from the optimized reactant to the final product are calculated from the coordinates collected from the 3D PES selecting a reaction path that passes through TS3.
 
 (i) \textbf{Coordinate 3} $(x)$ is the distance between the center of mass of NO$_{3}$ and the center of mass of NO multiplied by $\sqrt{\frac{\mu_{NO_{3}-NO}}{m_{O}}}$, where $\frac{1}{\mu_{NO_{3}-NO}} = \frac{1}{m_{N}+3*m_{O}} + \frac{1}{m_{N}+m_{O}}$

(ii) \textbf{Coordinate 4} $(y)$ is the bond length of NO multiplied by $\sqrt{\frac{\mu_{NO}}{m_{O}}}$, where $\mu_{NO}$ is the reduced mass of NO.
 
 When the distance between the center of mass of NO$_{3}$ and the center of mass of NO is large; the electronic energy is equal to the sum of the energy of NO$_{3}$ calculated in step 1 and the ground electronic energy of NO at equilibrium bond length. The potential energy curve from optimized reactant of step 2 to the point where the electronic energy is equal to the sum of the energy of NO$_{3}$ obtained in step 1 and the ground electronic energy of NO at equilibrium bond length is calculated by relax scan. For this part also mass weighted \textbf{Coordinate 3} and \textbf{Coordinate 4} are calculated. 


Like in Step 1, using \textbf{Coordinate 3} and \textbf{Coordinate 4}, new coordinate $r$ is calculated following the relation \ref{coord_2}
\begin{equation}
   r(i+1)=r(i) + \sqrt{(x(i+1)-x(i))^{2}+(y(i+1)-y(i))^{2}} \label{coord_2}
\end{equation}
where $r$(1)=0.

\begin{figure}
\centering
  \includegraphics[width=\linewidth]{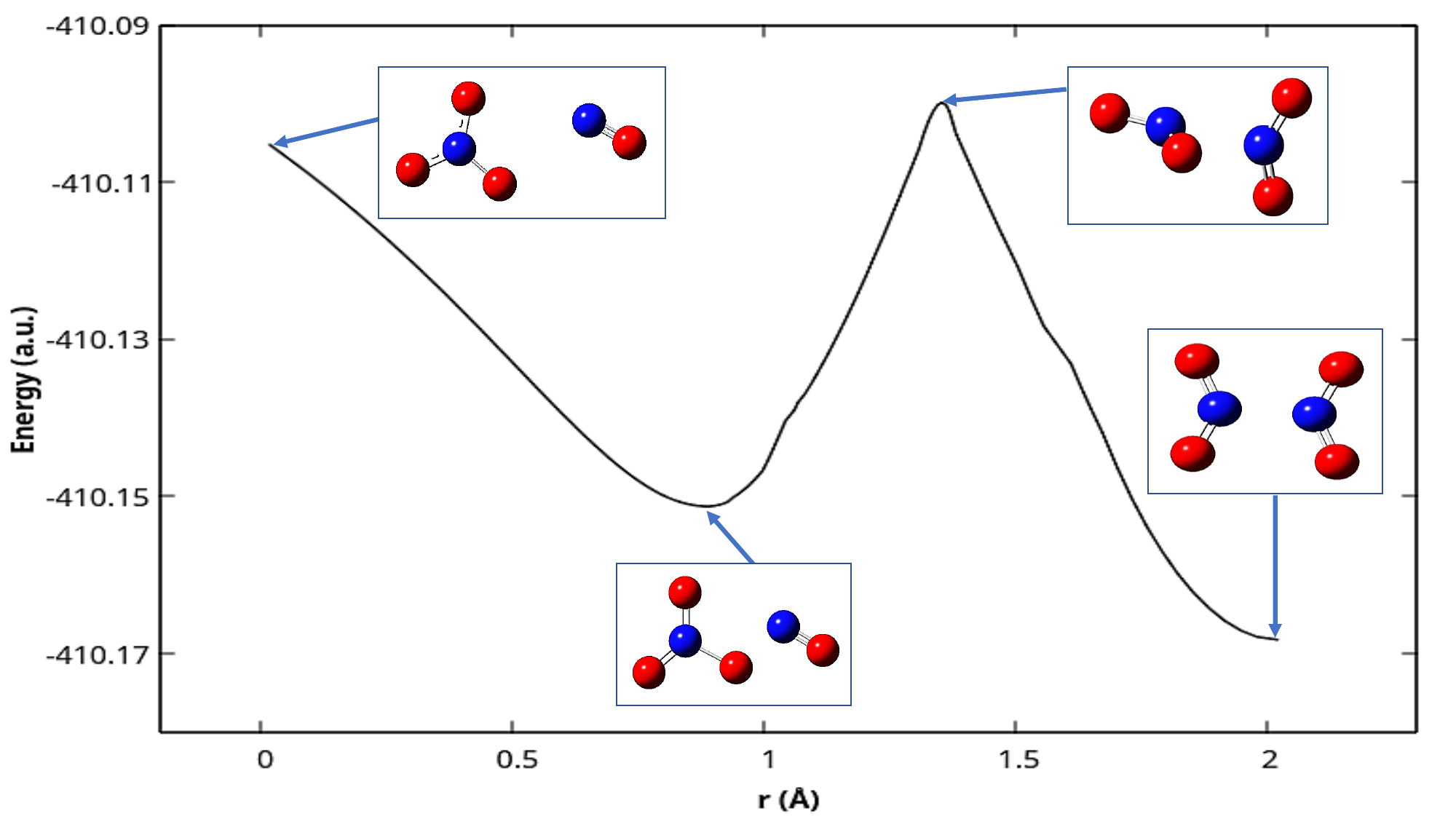}
  \caption{Potential Energy curve for step 2, NO$_{3}$+NO $\rightarrow$ 2NO$_{2}$ reaction along with optimized structures at stationary points. $r$ is the mass weighted coordinate obtained from coordinates 3 and 4 as defined in the text by using the relation given in equation \ref{coord_2}. The molecular structures, shown by arrows from left to right, correspond to the optimized structure at the point where the electronic energy is equal to the sum of the energy of NO$_{3}$ calculated in step 1 and the ground electronic energy of NO at equilibrium bond length, optimized structures of reactant, TS3 and  the final product (2NO$_{2}$).}
  \label{fig_07}
\end{figure}

 The potential energy curve obtained by using the new coordinate ($r$) is shown in Fig. \ref{fig_07} along with optimized structures.

Now, the ground electronic energy of NO at equilibrium bond length is added to the electronic energies of Step 1. Combining the potential energy curves of step 1 and step 2, we can obtain the complete potential energy curve of the complete reaction O$_{2}$+2NO $\rightarrow$ 2NO$_{2}$.

Following the adiabatic approximation of the spectrum, as explained in Ref. \cite{moiseyev2024conditions}, the reaction path Hamiltonian (RPH) is reduced into the following 1D effective Hamiltonian 
\begin{equation}
    H_{RC}=-\frac{1}{m_{O}}\frac{d^{2}}{dr^{2}} + V_{RC}(r)  \label{eq1}
\end{equation}
where $m_{O}$ is the mass of oxygen atom in atomic units. The complex poles that will be obtained by solving the Eq. \ref{eq1} are shape resonances that will be used to study the suppression of the reaction rate of the reaction \ref{r1}.
Here we should emphasis that the  Hamiltonian given in Eq. \ref{eq1} is NOT a 1D model but contains all information that is included in the multidimensional PES we have calculated where $V_{RC}(r)$ is the double barrier  potential along the \textit{one dimensional curve-linear reaction coordinate}.  See for example the calculations of the thermal reaction rate of Aziridine in Ref. \citenum{rom1993thermal} for a multidimensional potential energy surface for the price of one.

\subsection{Direct Calculation of Resonance Complex Poles of the Scattering Matrix}

Following our theoretical approach to suppress the reaction rate within the dark cavity, it is essential to ensure that the potential barrier supports complex poles. The direct calculation of the predissociation resonances of the scattering matrix is carried out by imposing outgoing boundary conditions on the solutions of the time-independent Hamiltonian of the $N_2O_4$ activated complex for the reaction $O_2+2NO\to 2NO_2$. The SES method is employed to impose the outgoing boundary conditions and to perform direct calculations of the complex resonance poles.

Notice that not all complex poles of the scattering matrix are associated with decay resonance states relevant for calculating the reaction rate. The so-called physical complex poles required for reaction rate calculations are those associated with outgoing boundary conditions that ensure the decay of the activated complex into products. Using the SES method, we calculate these physical resonance poles, which serve as the basis set in our computations of the reaction rates both outside and inside the cavity.
 
\begin{figure}
\centering
  \includegraphics[width=\linewidth]{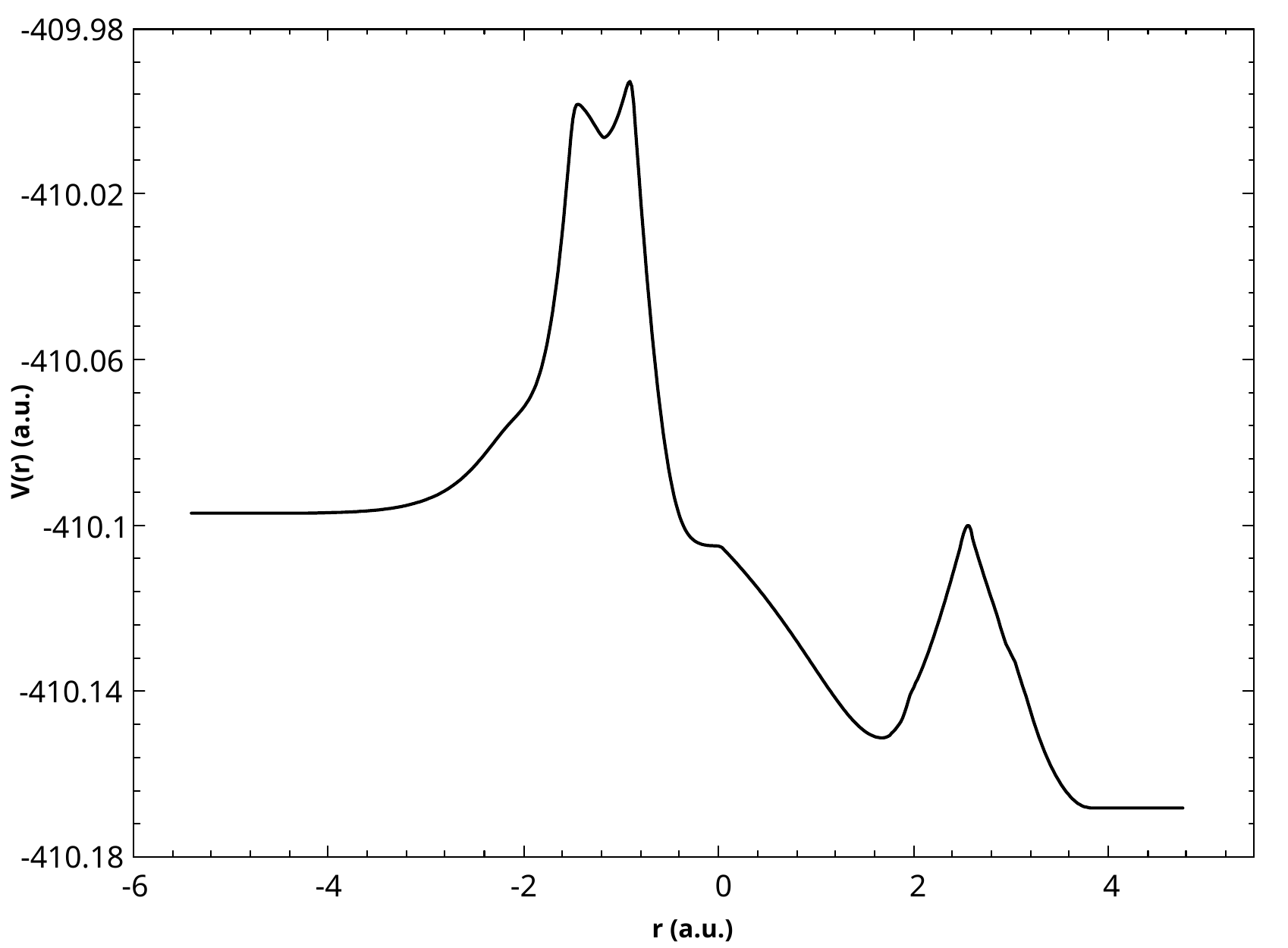}
  \caption{Complete Potential Energy curve as a function of the reaction coordinate r (a 1D curve linear coordinate in the multidimensional potential energy surface)}, after augmentation for the reaction O$_{2}$+2NO $\rightarrow$ 2NO$_{2}$.
  \label{fig_11}
\end{figure}

The potential energy surface consists of electronic energies which are calculated at equidistance grid points. The number of grid points are increased by 50 points on both sides by adding more points keeping the energy the same as the last coordinates. This augmentation is required while using SES as the scaling is done outside the interaction region in this method. From this point on, all calculations are done in atomic units. After augmentation, the final potential is given in Fig. \ref{fig_11}.

It is important to note that the 1D potential arises from the assumption that the potential energy barriers along the path from reactants to products are defined along the reaction coordinate, which is a curvilinear coordinate on the multidimensional potential energy surface. The calculations of the potential energy surface were carried out to study the reaction in the gas phase by solving the time-independent Schr\"odinger equation within the framework of the Born–Oppenheimer approximation. The temperature is assumed to be such that the relative kinetic energy of the reactants, $O_2$ and $2NO$, permits a classical transition just above the highest potential barrier for the conversion of reactants to products.

\subsubsection*{Smooth Exterior Scaling Method to impose outgoing boundary conditions to ensure the transition from reactants to products}

In SES\cite{moiseyev1998quantum,moiseyev2011non}, the Hamiltonian has the following form.

\begin{equation}
  \hat H\psi = E\psi
\end{equation}

where
\begin{equation}
\hat{H}=-\frac{\hbar^2}{2M}\frac{\partial^2}{\partial \rho^2} + V(\rho)
\end{equation}  
Here, $\rho=F(r)$ is a path in the complex coordinate plane which is defined as 
$\rho=F(r) \rightarrow re^{i\theta}$ where r is the reaction coordinate  $\rightarrow$ $\infty$.

The smooth exterior scaling path is defined as
\begin{equation}
 f(r)=\frac{\partial F}{\partial r}= 1+(e^{i\theta}-1)g(r) \label{ses2}.
\end{equation}

The value of $g(r)$ varies from 0 to 1 around the point $r=r_{0}$.

When $V(r\geq r_{0})=0$, then the unscaled potential $V(r)$ can be used instead of the complex potential $V(re^{i\theta})$. The effective Hamiltonian has the following form:
\begin{equation}
  \hat{H}=\frac{-\hbar^2}{2M}\frac{\partial^{2}}{\partial r^{2}} + V[F(r)] + \hat{V}_{CAP}
\end{equation}
\begin{equation}
  \hat H=\hat H_{0} + \hat V_{CAP}
\end{equation}
where $\hat{H_{0}}$ is physical Hamiltonian and

\begin{equation}
  \hat{V}_{CAP}=\hat{V}_{CAP}^{pot}+\hat{V}_{CAP}^{kin}
\end{equation}

$\hat{H_{0}}$ is physical Hamiltonian. $\hat{V}_{CAP}^{pot}$ and $\hat{V}_{CAP}^{kin}$ are kinetic and potential terms. The potential term $\hat{V}_{CAP}^{pot}$ can be ignored based on the conditions discussed in Ref. \cite{sajeev2006reflection}. Thus, 
\begin{equation}
  \hat{V}_{CAP}=\hat{V}_{CAP}^{kin}
\end{equation}
where
\begin{equation}
  \hat{V}_{CAP}^{kin}=V_{0}(r)+V_{1}(r)\frac{\partial}{\partial r} + V_{2}(r)\frac{\partial^{2}}{\partial r^{2}}  \label{vcap1}
\end{equation}
and
\begin{equation}
  V_{0}(r)=\frac{\hbar^{2}}{4Mf^{3}(r)}\frac{\partial^{2}f(r)}{\partial r^{2}}-\frac{5\hbar^{2}}{8Mf^{4}(r)}(\frac{\partial f(r)}{\partial r})^{2}
    \end{equation}
\begin{equation}
  V_{1}(r)=\frac{\hbar^2}{Mf^3(r)}\frac{\partial f(r)}{\partial r}
\end{equation}
\begin{equation}
  V_{2}(r)=\frac{\hbar^2}{2M}(1-f^{-2}(r))
\end{equation}

Now, g(r) in Eq. \ref{ses2} can be defined with the help of a specific family of integration paths in the complex coordinate plane as follows:

\begin{equation}
  g(r) = 1+0.5(tanh(\lambda (r-r_{ight}))-tanh(\lambda (r+r_{left}))) \label{ses3}
\end{equation}

By carrying out the integration over g(r), the complex path F(x) is obtained which has the following form-

\begin{equation}
    F(r)=r+(e^{i\theta}-1)[r+\frac{1}{2\lambda} ln\frac{cosh[\lambda(r-r_{right})]}{cosh[\lambda(r+r_{left})]}]
\end{equation}

Although the SES method depends on several parameters, $\lambda$, $\theta$, $r_{left}$, and $r_{right}$, each of the resonances is associated with a single cusp which is obtained when all parameters are held fixed and only $\theta$ is varied. In this type of calculation, which is a graphical method, each of the resonances is associated with a single curve where the absolute value of the velocity of the $\theta$-trajectory gets a minimal value\cite{moiseyev1981cusps}. The resonance energies thus obtained are stable with the variation of $\lambda$, $r_{left}$, and $r_{right}$.

\section{Results and Discussion}

\subsection{The suppression of the rate of the reaction O$_{2}$+NO $\rightarrow$ 2NO$_{2}$}

 In this work, the TS resonances of [ONOO]$^{\#}$ are accurately obtained from the 1D effective potential barrier that is obtained within the framework of the nuclear adiabatic approximation. In this case, all complex poles are associated with the potential barrier and are physical poles and the rate of the reaction is suppressed by the dark cavity that couples the TS resonance with one of the shape resonances of ONOO. The values of the parameters used in this calculation are $\lambda$=2.0 a.u. and r$_{right}$=4.35 a.u. and r$_{left}$=4.73 a.u. The corresponding resonance wave functions are calculated at $\theta$=0.14 radian which is the stationary point of the resonance $\theta$-trajectories.

 Formally, the absolute energy of the complex excited resonance state that is associated with TS resonance should be equal to the energy of the top of the potential energy plus the zero vibrational energy. However, in our calculations, we do not include the vibrations, which are perpendicular reaction coordinates. We limit our calculations to the zero-order perturbation Hamiltonian where we used as potential the electronic energy as a function of the reaction coordinate. The real part of the complex eigenvalue is about equal to the energy of the top of the potential energy barrier, which is taken as the TS resonance. We choose the TS resonance as the complex pole located at the top of the potential barrier, as it describes the reaction occurring outside the cavity without considering quantum tunneling, and is defined according to standard transition state theory.

\begin{figure}
\centering
  \includegraphics[width=\linewidth]{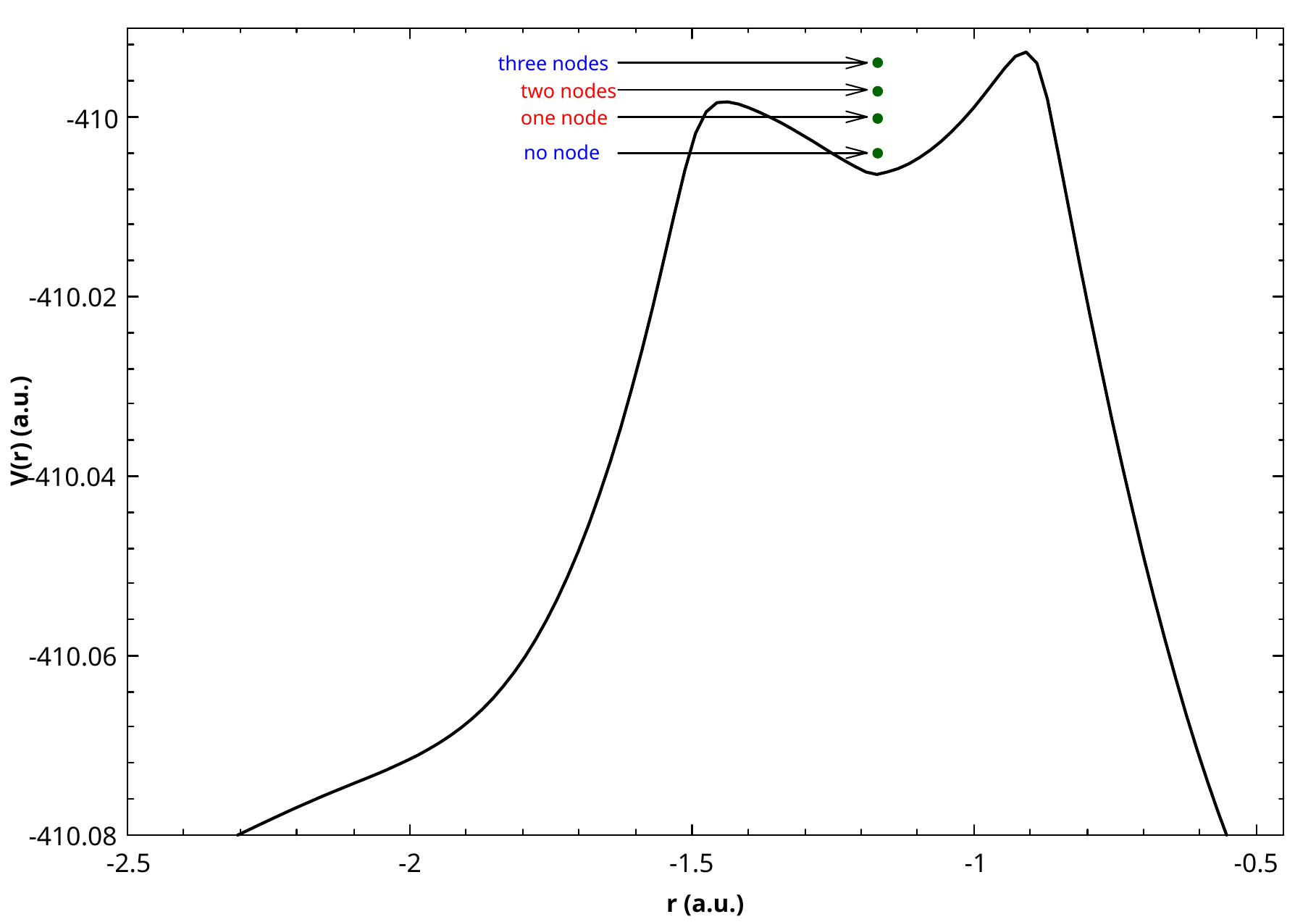}
  \caption{A zoom of potential barrier given in Fig. \ref{fig_11} showing the positions of the shape type resonances assigned by the number of the nodes. The corresponding resonance widths are given in table \ref{Table_1}. The three nodes resonance is taken as the transition state (TS) resonance of the reaction as determined the rate of the reaction outside of the cavity (see the explanation in the text). The photon energy in the cavity is taken as the energy difference between the three-node and zero-node resonances.}
  \label{fig_p1}
\end{figure}

\begin{figure}
\centering
  \includegraphics[width=\linewidth]{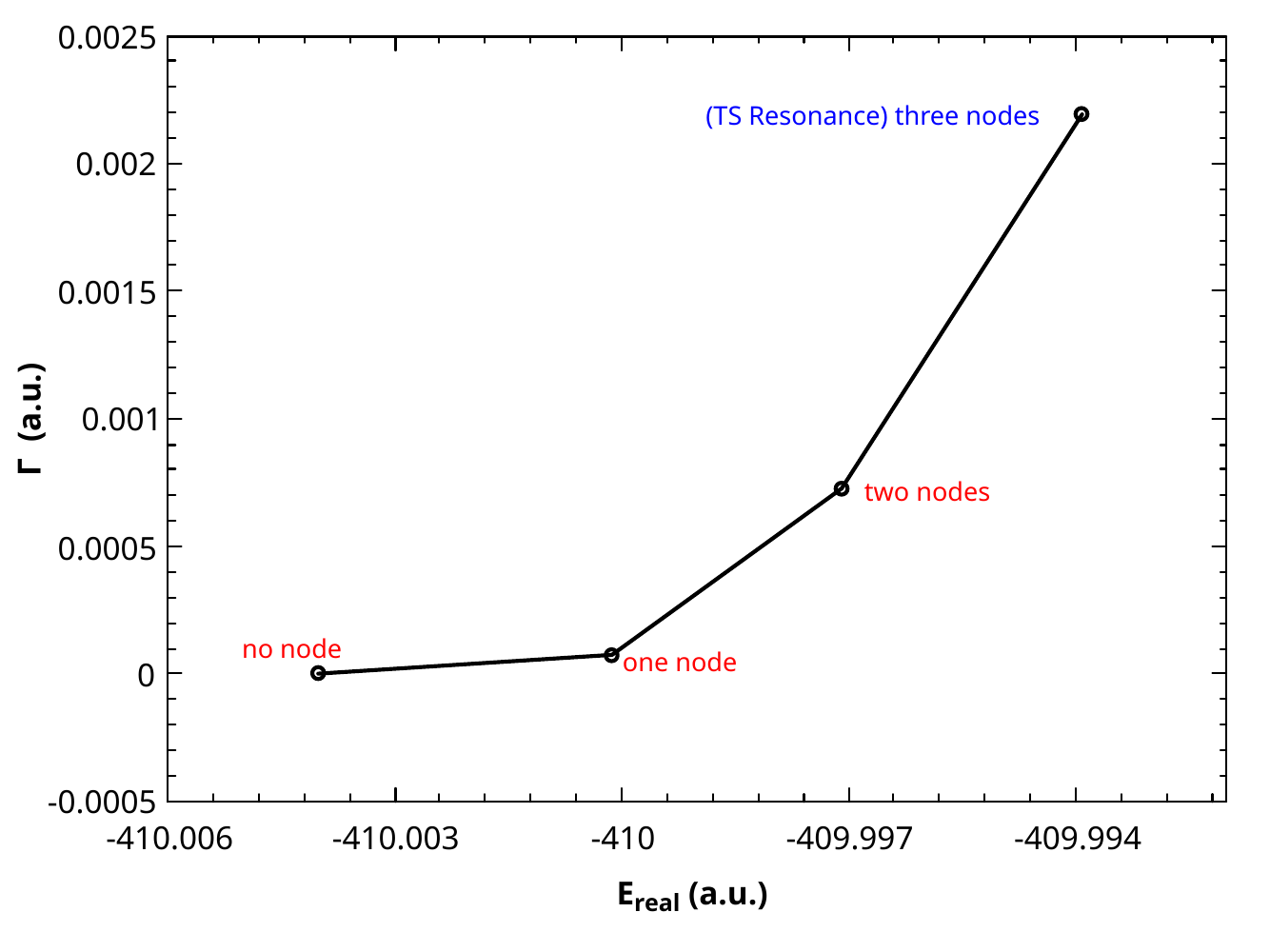}
  \caption{The lowest four resonance complex poles calculated by SES method. The number of nodes are counted from the plots of the square integrable complex scaled real part of $\psi$ (shown in the SI). The transition state is associated with the three node resonance with shortest lifetime which is located at the top of the potential barrier as explained in the text.}
  \label{fig_p2}
\end{figure}

In Fig \ref{fig_p1}, the potential barrier for the reaction \ref{r1} and the positions of TS and other shape resonances are shown. The potential barrier is asymmetric. The resonance positions were calculated using the SES method\cite{moiseyev1998quantum,moiseyev2011non}. The resonance with three nodes represents the transition state (TS) resonance since it is embedded at the top of the largest potential energy surface (see supplementary information). As explained by Seideman and Miller, its imaginary part is associated with the chemical reaction rate, as obtained from semiclassical calculations\cite{seideman1991transition}. As will be explained later, the cavity parameters are chosen such that the three-node resonance (associated with the transition state of the reaction from reactants to products) is coupled to the two-node resonance via the emission of a single photon, achieved by properly selecting the distance between the cavity mirrors. Furthermore, the transition state can also couple to the no-node resonance through the emission of multiple photons when ten molecules interact simultaneously with the cavity.

In Fig. \ref{fig_p2} the predissociation resonances of the activated complex in the $O_2+2NO\to 2NO_2$ reaction are presented along with the corresponding number of nodes. The TS resonance is marked by blue color. The TS resonance is localized at the top of the potential barrier and is associated with the shortest lifetime resonance.

The ability to suppress the rate of the reaction 2NO + O$_{2} \rightarrow$ 2NO$_{2}$ is due to the presence of resonance that has less than three nodes. The no-node resonance  is chosen to be coupled to the TS resonance by the cavity since it has the lowest resonance width (smallest decay rate). Notice that the TS resonance emitting one photon (induced by the interaction of the molecules with the cavity) will have the same energy as the no node resonance. However, when several (up to ten) molecules are inside the cavity and simultaneously interact with it at the transition state (TS) configuration, the cavity frequency is tuned to couple the TS with the two-node resonance.

The following table is prepared using the real and imaginary parts of the above resonances.

\begin{table*}
\caption{\label{tab:table3}Resonance energies and widths calculated by SES method with corresponding complex dipole transitions. The cavity frequency and wavelengths calculated from the energy difference between j$^{th}$ and ground state resonance (j=0).}
\begin{ruledtabular}
\begin{tabular}{ccccc}
 Node&E$^r$ (a.u.)&$\Gamma$ (a.u.)&$\hbar\omega_{cav}$ (a.u.)
&$\lambda$ ($\mu$m)\\ \hline
 0 & -410.004009 & 1.06$\times 10^{-6}$ & - & - \\
 1 &	-410.000126 & 7.46$\times 10^{-5}$ & 0.0038830 & 11.7282195\\
 2 & -409.997093 & 7.26$\times 10^{-4}$ & 0.0069161 & 6.5846364\\
 3 & -409.993916 & 0.00219138 & 0.0100931 & 4.5120284\\
 
\end{tabular}
\end{ruledtabular}
\label{Table_1}
\end{table*}



  

The symbols in the tables are defined below.

\begin{eqnarray}
&& E_i^r=Real[E_i^{res}] \nonumber \\ &&
    \Gamma_i=-2Im[E_{i}^{res}]
\end{eqnarray}

   $\omega_{cav}$ is the cavity frequency and is the energy gap between the { TS resonance (3-nodes in our case) and the ground state resonance (GR) with 0-node.} Namely,  $\hbar\omega_{cav}=E^r_{\color{blue}3} - E^r_0=Real[E_{res}^{TS}-E_{res}^{GR}]$ and the distance between two cavity mirrors is given by, $L = \frac{\lambda}{2}$ where $\lambda=\frac{2\pi c}{\omega_{cav}}$. The dipole transitions from one predissociation resonance of the activated complex to another is given by

\begin{equation*}
d_{i,j}=\frac{(\psi^{res}_{i}|F(r,\theta)|\psi^{res}_{j})}{\sqrt{(\psi^{res}_{i}|\psi^{res}_{i})(\psi^{res}_{j}|\psi^{res}_{j})}} 
\end{equation*}

where 
$F(r,\theta)$ is the smooth exterior scaling (SES) contour of integration, which ensures the imposition of outgoing boundary conditions on the solutions of the time-independent Schr\"odinger equation for the potential governing the reaction  $O_2+2NO\to 2NO_2$.

\subsection{The cavity-coupled polaritionic Hamiltonian and the decay rate or $O_2+2NO\to 2NO_2$ inside of the dark cavity}

In polaritonic chemistry, the cavity‑coupled Hamiltonian describes how molecular degrees of freedom interact with the quantized electromagnetic modes of an optical cavity.
The Hamiltonian of a molecule inside a cavity in the Length Gauge representation is  given in Ref. \cite{rokaj2018light}. Here, we neglect the self energy term  and the effective Hamiltonian that describes one molecule out of the ensemble of N molecules  that interact with a one photon cavity is given by,

\begin{widetext}
\[H^{pol}(\varepsilon_{cav},\omega_{cav})=
\begin{bmatrix}
E_{0} & 0 & 0 & 0 & 0 & d_{01}\varepsilon_{cav} & d_{02}\varepsilon_{cav} & d_{03}\varepsilon_{cav} \\
0 & E_{1} & 0 & 0 & d_{10}\varepsilon_{cav} & 0 & d_{12}\varepsilon_{cav} & d_{13}\varepsilon_{cav} \\
0 & 0 & E_{2} & 0 & d_{20}\varepsilon_{cav} & d_{21}\varepsilon_{cav} & 0 & d_{23}\varepsilon_{cav} \\
0 & 0 & 0 & E_{3} & d_{30}\varepsilon_{cav} & d_{31}\varepsilon_{cav} & d_{32}\varepsilon_{cav} & 0 \\
0 & d_{10}\varepsilon_{cav} & d_{20}\varepsilon_{cav} & d_{30}\varepsilon_{cav} & E_{0}-\hbar\omega_{03} & 0 & 0 & 0 \\
d_{01}\varepsilon_{cav} & 0 & d_{21}\varepsilon_{cav} & d_{31}\varepsilon_{cav} & 0 & E_{1}-\hbar\omega_{03} & 0 & 0 \\
d_{02}\varepsilon_{cav} & d_{12}\varepsilon_{cav} & 0 & d_{32}\varepsilon_{cav} & 0 & 0 & E_{2}-\hbar\omega_{03} & 0 \\
d_{03}\varepsilon_{cav} & d_{13}\varepsilon_{cav} & d_{23}\varepsilon_{cav} &0 & 0 & 0 & 0 & E_{3}-\hbar\omega_{03}
\end{bmatrix} 
\]
\end{widetext}

where the polarization direction is taken to be tangential to the transition state along the reaction
coordinate and $E_0,E_1,E_2,E_3$ are the vibrational states of the field free resonances for the  reaction $O_2+2NO\to 2NO_2$ which are counted by the number of their nodes  (see Fig.4). $\varepsilon_{cav}$ is the strength of the coupling between every of the molecules and the cavity which is reduced to zero as the distance between two cavity mirror becomes very large (i.e., $L \rightarrow \infty$). Note that the cavity frequency is taken here as the energy gap between the energy of the three-node resonance (the TS resonance) and the lower-energy vibrational level of the zero-node resonance. We use a generalization of the Jaynes-Cummings model (Ref. \cite{jaynes2005comparison}) by not using the rotating wave approximation. Therefore, in principle we could take into consideration many photon interactions. Since the cavity frequency is taken here as $\omega_{03}$ (i.e., under the resonance condition), we limit ourselves to single-photon interactions, where the permanent dipole terms are set to zero, and only dipole transitions accompanied by the emission or absorption of a single photon are allowed.
Here, the couplings between the TS resonance and all other resonances are enabled. The  following 8$\times$8 matrix of the polariton Hamiltonian is prepared where the basis set consists of the four shape-type resonances presented in Fig. 5 in the vacuum and the four resonances in the $\vert$-1$_{photon}\rangle$ QED state, implying the emission of one photon into the cavity.  {The minus sign is to remind the reader that this photon  is formed by the TS that emit one photon being inside the cavity.}

\begin{figure}
\centering
  \includegraphics[width=\linewidth]{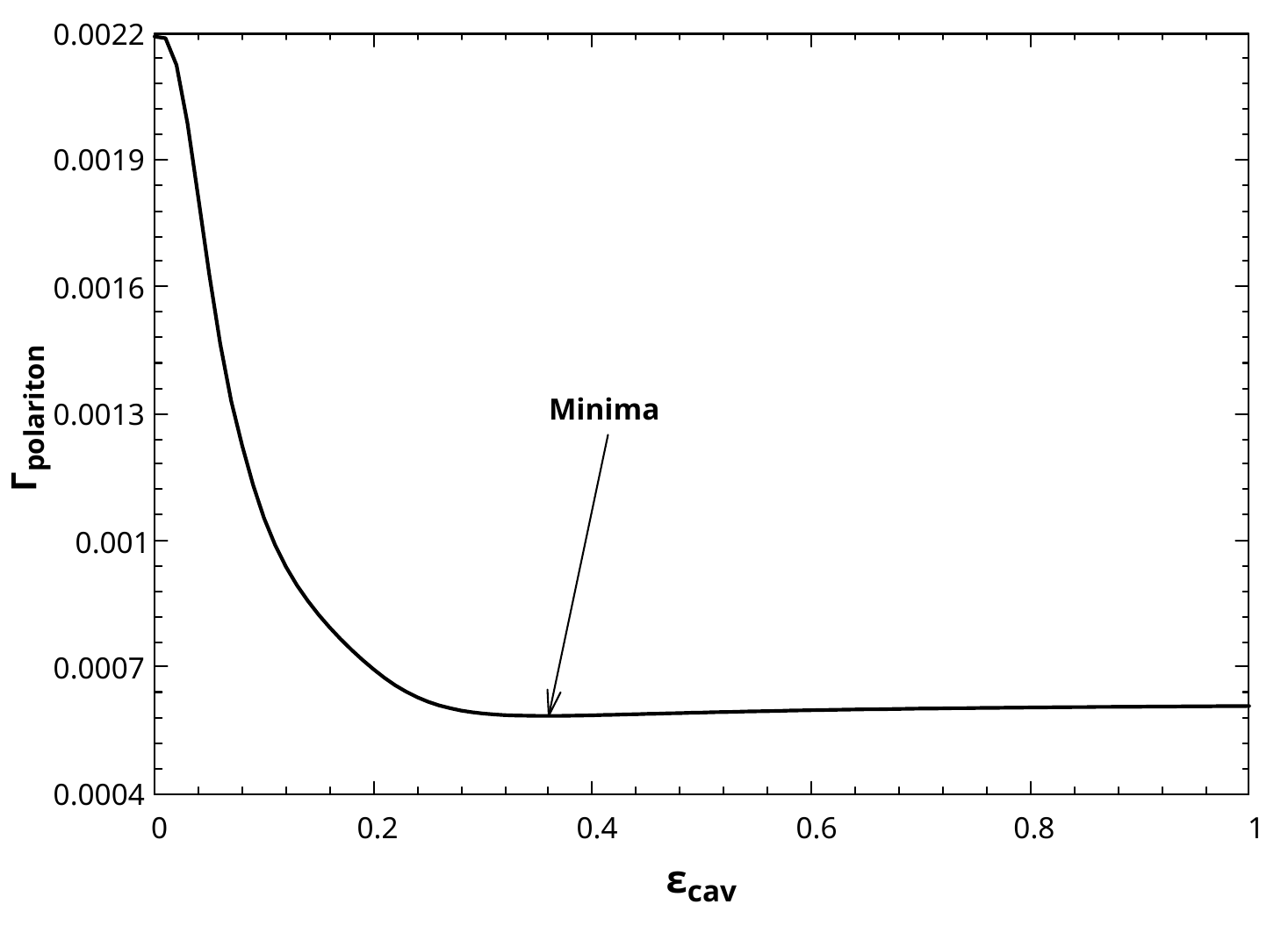}
  \caption{The effect of the cavity on the reaction rate, $\Gamma_{polariton}$, of the TS polaritons in the reaction O$_{2}$ + 2NO $\rightarrow$ 2NO$_{2}$ is shown while the couplings between the TS resonance and all other shape-type resonances are enabled. This figure demonstrates that when all resonances are included in the basis set calculations, the reaction rate is reduced by a factor of 3 (the rate of reaction obtained for $\epsilon_{cav}=0$ is the rate of reaction outside of the cavity). All units are given in atomic unites.}
  \label{pol08}
\end{figure}

In this case, the reaction rate, $\Gamma_{polariton}$ can be calculated from the following relation

\begin{equation}
    \Gamma_{polariton}=\sum_{\alpha=1}^{8}\Gamma_{\alpha}|C_{8\alpha}|^{2}
\end{equation}

by solving the TISE
\begin{equation}
   H^{pol}(\varepsilon_{cav},\omega_{cav})C_{\alpha} = (E^{r}_{\alpha}-i\frac{\Gamma_{\alpha}}{2})C_{\alpha}
\end{equation}
for every eigenvector $\alpha$=1,2,...,8 and  $\sum_{j=1}^{8} |C_{j\alpha}|^{2}=1$, where $C_{j\alpha}$ is the j$^{th}$ element of $C_{\alpha}$ vector. The $C_{8\alpha}$ is the projection of the $\alpha-th$  polaritonic energy on the $|TS\rangle$ function associated with the $E^r_{TS}-  \hbar\omega_{cav}$ field free  energy. The reaction rate, $\Gamma_{polariton}$ as a function of $\varepsilon_{cav}$ is plotted in Fig. \ref{pol08}. The results presented in the figure clearly show that the reaction rate is reduced by factor 3.

This is a single-molecule reaction study. However, experimental work on cavity-modified chemistry typically involves ensembles of millions of molecules, where collective effects and the role of dark states are central.  See, for example, Ref. \cite{li2022molecular}. On the basis of the proof given in this review only one bright state is coupled to the cavity with the strength coupling of $\epsilon_{cav}=\epsilon_{QED} \sqrt{N}$ when $\epsilon_{QED}$ is about $10^{-8}$ a.u. Since the number of the dark states is $N-1$, one might not expect to measure the cavity effect on the rate of the reaction inside the cavity since all the reactions of the  dark states remains as it in free space (outside of the cavity).

However, our situation is fundamentally different. In the scenarios studied in the review by Nitzan and his coauthors, all N molecules initially reside in their ground states, while a single photon is already present in the cavity. This photon interacts coherently with a linear superposition of all possible single-molecule excitations within the ensemble, resulting in collective polaritonic dynamics.

In contrast, in our case, all N molecules are initially in an excited transition state (TS) and the cavity is initially in the vacuum state. These excited molecules can relax to a lower excited state via photon emission, corresponding to all reactants having collided at short range and reached the TS. The number of photons that may be emitted ranges from one to N, depending on the relaxation pathways.

\begin{figure}[h]
\centering
  \includegraphics[width=\linewidth]{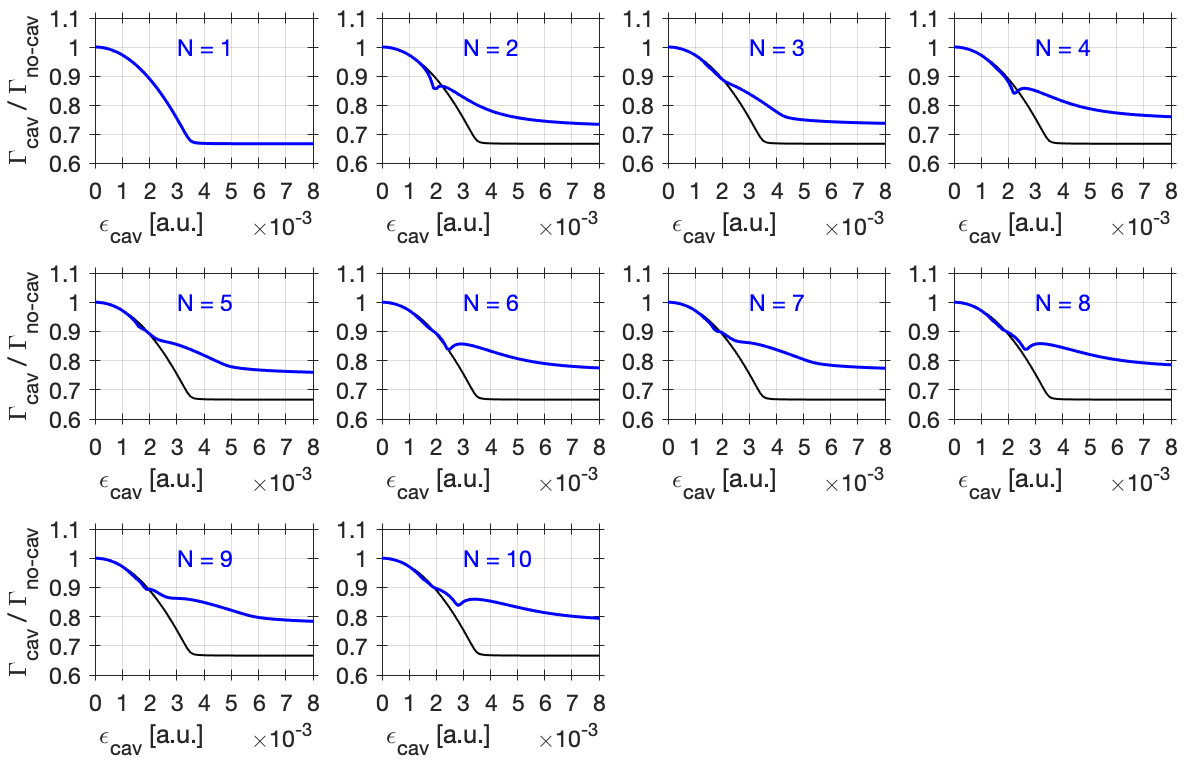}
  \caption{Calculated results for the suppression of the reaction rate due to the initially dark cavity for the reaction 2NO+O$_2$ (see details in the text). Shown (in blue lines) is the cavity-induced rate suppression, R$_{cav,no-cav}$ = $\Gamma_{cav}/\Gamma_{no-cav}$, as a function of the coupling parameter between the cavity and a single molecule, $\varepsilon_{cav}$, for different numbers of molecules in the cavity, N . The results for N = 1 are shown, for comparison, in all the panels of the figure (solid black lines). The quantities $\Gamma_{cav}$ and $\Gamma_{no-cav}$ denote the reaction rates with and without the cavity (i.e, inside and outside the cavity), respectively. See details in the text}
  \label{pol00}
\end{figure}

Consequently, the effect of the cavity on the reaction rate is not necessarily observable in the same way as in polaritonic chemistry. See the Supplementary Information in Ref.\cite{berenstein2025} 
 for the use of the Travis-Cummings model\cite{tavis1968exact} when the basis set are the polaritonic states given by $$\{|n_{photon}\rangle|\chi(GR,n_{photon};TS,N-n_{photon})\rangle\}_{n_{photon}=0,1,..,N},$$where $\chi(GR,n_{photon};TS,N-n_{photon})\rangle$ is a function of all possible combinations with $N_{photon}$-molecules in GR resonance states and $(N-N_{photon})$-molecules in the TS resonance state. See the description of the polariton basis set given in the introduction. Using this polariton basis set we diagonalize the following Hamiltonian that includes the indirect interactions between the molecules inside the cavity via the cavity:
 \begin{equation}
 \label{HAM-N-CAV}
     \hat H_{mol-cav}(N)=\Sigma_{j=1}^N [\hat H_{j-mol} + \hat H_{j-mol/QED}]
 \end{equation} 
 where,

\begin{widetext}
    
 \begin{equation}
\hat H_{j-mol}=(E^r_3-\frac{i}{2}\Gamma_3-\hbar\omega_{cav})|TS\rangle_{j-mol}\langle TS|_{j-mol}+(E^r_0-\frac{i}{2}\Gamma_0)|GR\rangle_{j-mol}\langle GR|_{{j-mol}}
 \end{equation}
 and
 \begin{eqnarray}
&&\hat H_{j-mol/QED}=\epsilon_{cav}d_{TS,GR}\Big(|TS\rangle_{j-mol}\langle GR|_{j-mol} +|GR\rangle_{j-mol}\langle TS|_{j-mol}\Big) \nonumber \\ && 
\cdot\Big(|1_{photon}\rangle_{j-mol}\langle 0_{photon}|_{j-mol}+|0_{photon}\rangle_{j-mol}\langle 1_{photon}|_{j-mol}\Big)
 \end{eqnarray}

 \end{widetext}
 
 where $d_{TS,GR}$ is the dipole transition between the 3-node and 0-node resonances as shown in Fig. 5, {Assuming one resonance condition, where the TS due to the emission of one photon has the same energy as the GR state and therefore they are most strongly coupled inside the cavity, when $\epsilon_{cav}$ is the coupling strength parameter between a single molecule and the cavity.} By diagonalization of the Hamiltonion given in Eq. \ref{HAM-N-CAV} using the polaritonic basis functions as defined in the introduction the polariton complex energies of N molecules inside the cavity, $E_{N-mol-in-cavity}^{\alpha}(\epsilon_{cav})$, and the associated polariton wavefunctions, $|N-mol-in-cavity(\epsilon_{cav})\rangle^{\alpha}$ are obtained.
 The rate of the reaction inside the cavity is given by,

 \begin{widetext}
    
 \begin{equation}
\Gamma_{cav}(\epsilon_{cav})= -2\Sigma_{\alpha}|\langle Polariton|_{0-photon}|N-mol-in-cavity(\epsilon_{cav})\rangle|^2
Im E^{\alpha}_{N-mol-in-cavity}(\epsilon_{cav})
 \end{equation}
  \end{widetext}
 
 while the reaction rate outside of the cavity is given by,
 \begin{equation}
 \Gamma_{no-cav}=\Gamma(\epsilon_{cav}=0)
 \end{equation}
 In Fig. \ref{pol00} we present the ratio $\Gamma_{cav}(\epsilon_{cav})/\Gamma_{no-cav}$ as a function of $\epsilon_{cav}$
  for the $O_2+2NO\to 2NO_2$ reaction.

As one can see, the ratio between the rates inside and outside of the cavity shows that the  suppression of the rate for one molecule and 10 molecules are almost identical for a relatively weak interactions of the molecules with the cavity. Both show suppression of 10 to 15 percent which are hopefully measurable. Notice that in gas phase reactions the number of molecules inside the cavity can be millions, but only a small fraction of them can be simultaneously at the transition state configuration of the activated complex. Due to the computational difficulty of diagonalizing matrices larger than 1024 × 1024, we limit the number of molecules in the cavity—each interacting identically with the cavity mode—to a maximum of 10.

Moreover, in this work, we focus on the new mechanism presented, which can be verified experimentally by measuring the relative kinetic, electronic, and vibrational energies of the reactants and products. According to the proposed mechanism, in a dark cavity { (i.e., no photon is inserted into the cavity)}, the total energy of the products will be lower than that of the reactants, with the difference corresponding to the energy of the emitted photons. Specifically, inside the cavity, the reaction proceeds as:
$O_2+2NO\to 2NO_2+\hbar\omega_{cav}$
whereas outside the cavity, no such photon emission occurs and the reaction proceeds as:
$O_2+2NO\to 2NO_2$.

\section{Summary and Concluding Remarks}\label{sec13}
In this work we provide the conditions for suppression of rate of reactions by dark cavity that consists of two parallel mirrors. The distance between the two mirrors and the selecting the appropriate chemical reactions that its rate can be suppressed are given here. The potential energy surface of N$_{2}$O$_{4}$ was calculated and the reaction coordinate was determined by using the mass weighted coordinates for the potential energy surface. The double well potential along the reaction coordinate supports shape type resonances which are calculated by the use of the SES approach. The resonance with three nodes is associated with the transition state resonance that determine the reaction rate of $\frac{1}{2}$O$_{2}$+NO $\rightarrow$ NO$_{2}$ of the reaction outside of the cavity. The cavity couples the TS resonance with  all the lower energy resonances  than the energy of the TS and thereby reduces the chemical rate. The suppression of the reaction rate depends on the distance between the mirrors, which determines the cavity parameters, and on the number of molecules in the transition state configuration that simultaneously interact with the cavity. We propose to measure the total energy of reactants, $O_2+2NO$, and compare it with the total energy of the products $2NO_2$. We expect on the basis of our calculations that the total energy of the reactants (including relative kinetic energy) will be smaller than the total energy of the products (including relative kinetic energy) by the photon energy $\hbar\omega_{cav}$.

\begin{acknowledgments}
Israel Science Foundation (ISF) grant No. 1757/24 is acknowledged by Mwdansar Banuary and Nimrod Moiseyev for a partial support. Prof. Zohar Amitay from the Schulich Faculty of Chemistry at the Technion is acknowledged for providing us the Fig.7.
\end{acknowledgments}

\section*{Data Availability Statement}

The data that support the findings of this study are available within the article [and its supplementary material].

\nocite{*}
\bibliography{aipsamp}

\end{document}